**A novel mathematical technique to assess of the mitral valve dynamics based on echocardiography**

Mersedeh Karvandi [1] MD, Saeed Ranjbar [2,*] Ph.D, Seyed Ahmad Hassantash MD [2], Mahnoosh Foroughi MD [2]

**Research institute:**
1- Taleghani Hospital, Shahid Beheshti University of Medical Sciences,Tehran, Iran
2- Institute of Cardiovascular Research, Modarres Hospital, Shahid Beheshti University of Medical Sciences, Tehran, IR Iran

***Corresponding author:**
Saeed Ranjbar
Institute of Cardiovascular Research, Modarres Hospital, Shahid Beheshti University of Medical Sciences, Tehran, IR Iran
E-mail: sranjbar@ipm.ir
Tel/FAX: +9821 22083106

## Abstract:

**Purpose:**

The mechanics of the mitral valve leaflet as a nonlinear, inelastic and anisotropic soft tissue results from an integrated response of many mathematical/physical indexes' that illustrate the tissue. In the past decade, finite element modeling of complete heart valves has greatly aided evaluation of heart valve surgery, design of bioprosthetic valve replacements, and general understanding of healthy and abnormal cardiac function. Such a model must be based on an accurate description of the mechanical behavior of the valve material. It is essential to calculate velocity/displacement and strain rate/strain at a component level that is to work at the cellular level. In this study we developed the first three-dimensional displacement vectors field in the characterization of mitral valve leaflets in continuum equations of inelasticity framework based on echocardiography.

**Method:**

Much of our knowledge of abnormal mitral valve function is based on surgical and post-mortem studies while these studies are quantitative in some cases, they are limited by evaluation of valve anatomy in a fixed and nonfunctioning state. A more sophisticated analysis method is necessary to gain a full considerate of mitral valve function. Several groups attempted to model mitral valve anatomy and function by mathematical/physical equations.

**Result:**

Preliminary results concerning a different aspect of MVL biomechanics, such as leaflets dynamics, displacements/velocities and strain rates/strains of points on leaflets, were in good agreement with in echocardiographic observations.

**Conclusion:**

Quantitative information on MVL and dynamics can be extracted from equations of inelasticity, when performed in multiple TTE and TEE planes. These data potentially allow the implementation of an image-based approach for patient-specific modeling of MVL. This advance could overcome the limitations of previously proposed models and give new insight into the complex MV function. This approach could constitute the basis for accurate evaluation of MV pathologic conditions and for the planning of surgical procedures.

**Keyboards:** Mitral valve leaflets, Mathematical modeling, Echocardiography

**Introduction:**

The left side of the heart accepts oxygenated blood at low pressure from the lungs into the left atrium. The blood then moves to the left ventricle which pumps it forward to the aorta to circulate the body. The heart's valves maintain the unidirectional flow of the blood through the heart, e.g. the mitral valve prevents regurgitation of blood from the left ventricle into the left atrium during ventricular systole and the aortic valve prevents blood flowing back from the aorta into the left ventricle during diastole. The principal fluid phenomena involved in the left ventricular diastolic flow are related to the presence of symmetric structures that develop with the strong jet that enters through the mitral valve (Reul et al., 1981; Saber et al., 2001) [1,2]. It is conjectured that the fibered structure has important effects on the function of the ventricle (Bacanni et al., 2003; Daebritz et al., 2003; Pierrakos and Vlachos, 2010) [3-5].

The human mitral valve is a complex anatomical structure consisting of two valve leaflets, an annulus, chordate tendineae, and two papillary muscles which are finger-like projections embedded into the underlying left ventricular myocardium. The mitral annulus is a saddle shaped fibrous ring which seamlessly transitions into the two leaflets (Grashow et al., 2006; van Rijk-Zwikker et al.,1994) [6,7] Figure 1.The leaflets extend into the left ventricle where they are tethered to the papillary muscles via an intricate arrangement of chordae tendineae. The chordae tendineae consist of a complex web of chords that attach all over the leaflets of the valve. They prevent the prolapse of the valve leaflets at systole, and additionally assist in maintaining the geometry and functionality of the ventricle.

The papillary muscles play an important role and are believed to lengthen during isovolumetric contraction and shorten during ejection as well as during isovolumetric relaxation to maintain the chordae at the same deformation level, during opening and closure (Marzilli et al., 1980) [8].

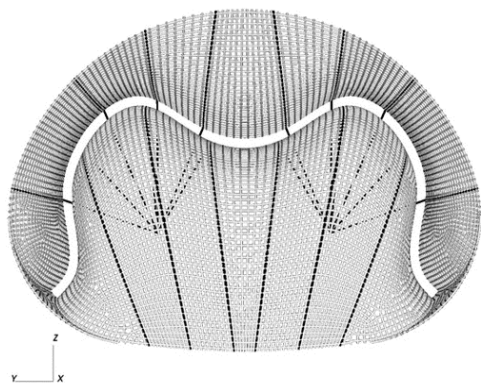

Figure 1: The presentation of the non-planar, annular saddle-shaped of the mitral valve

The Mitral valve structure and function is one of the parts in providing/completing for the velocity vector field attached to mitral valve leaflets. Hence, the geometrical Mitral valve parameters including: 1- Anterior tending angel, 2- Anterior leaflet tending angle. 3- Posterior leaflet tending angle, 4- Anterior bending angle, 5- Length of leaflet coaptation. 6- Tending height, 7- Tending area, 8- The length of leaflets, 9- the circumference of the annulus of the mitral valve, 10- Commissure distance, 11- Systolic Antero-Poster distance. 12- Diastolic Antero-Poster distance and13- Annular height have not been considered at the recent studies in mitral valve leaflets modeling. These measurable echocardiographic data as the elementary data are at least needed and the main goal of this paper to reconstruct a mitral valve model to providing a boundary condition of the fluid dynamic of the mitral valve leaflets Figure 2.

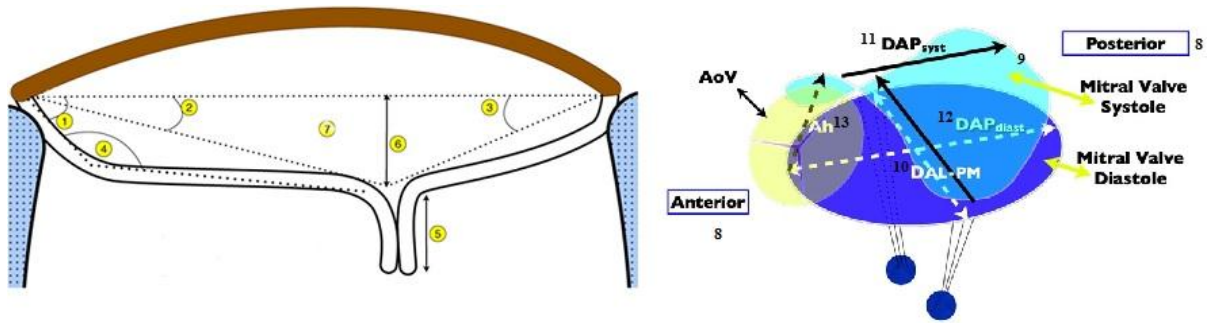

Figure 2: These 13 Parameters used to define the geometry of Mitral valve leaflets. AoV, aortic valve; Ah, annular height; DAP, anterior-posterior diameter; Syst, systole; Diast, diastole;

Method and Result:

Echocardiography imaging of the mitral valve motion during the cardiac cycle of planes TTE in the short axis, long axis and four chamber views, and multi-planes TEE in the lower esophageal views were prospectively acquired for 200 patients of 65 time-frames from diastole to systole (early diastole, mid diastole, atrial systole, end of diastole, end of systole). In each plane and for each frame during diastole to systole, 13 stated geometrical parameters of mitral valve annulus and leaflets were automatically measured by utilizing TomTec software (Philips Medical, Andover, MA) Figure 3.

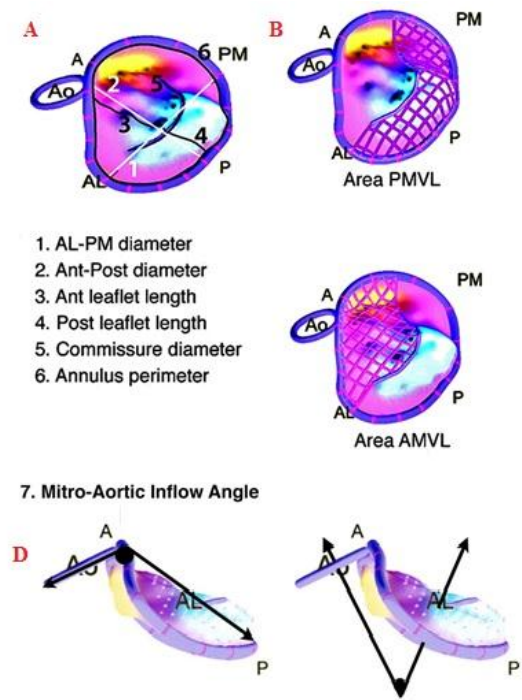

**Figure 3:** The images show the quantitative evaluation of the mitral valve, which is shown with TomTec software. The images show the marking of the leaflets, which is done in several points of the cardiac cycle. The images in A, B, and C are generated from the data set. From these, measures of annular dimensions, leaflet areas/lengths, and relations are obtained. The latter (C) include measure of the mitro-aortic angle (left: inflow; right: outflow) to help predict risk of systolic anterior motion of the mitral leaflets. AL, anterolateral; PM, posteromedial; LA, left atrium; LV, left ventricle; Ao, aortic valve; A, anterior; P, posterior; PMVL, posterior mitral valve; AMVL, anterior mitral valve leaflet

When those 13 parameters are assessed then vectors/lines that attach bases, middles and tips of AML and PML together, were manually identified using MATLAB software in the acquired 65 time-frames from diastole to systole Figure 4.

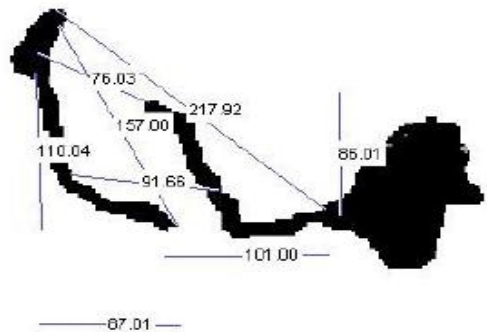

**Figure 4:** Gray lines provide vectors that attach bases, middles and tips of AML and PML together with MATLAB software.

A 3D AML and PML positions were automatically computed for each frame and used as input to displacement vectors modeling of the mitral valve leaflets Figure 5.

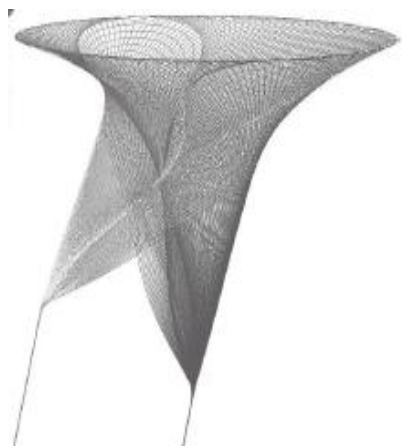

Figure 5: 3D AML and PML positions.

By solving mathematical equations of inelastic properties of leaflets, each plane was replaced with 60 vectors of displacement and ultimately the mitral valve leaflets were realized by 1125 vectors Figure 6 of displacements where show also translations, rotations and pure strains of bases, middles and tips of AML and PML simultaneously per cardiac cycle [9-14].

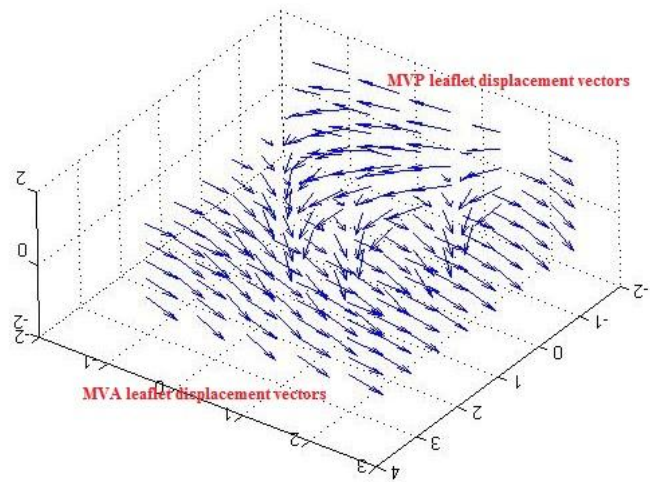

**Figure 6**: Displacement vector assignments to mitral valve leaflets

## Conclusion:

Resulted data potentially allow the implementation of an image-based approach for patient-specific modeling of mitral valve leaflets. This approach could constitute the basis for accurate evaluation of mitral valve pathologic conditions and for the planning of surgical approaches.

**Conflict of interest:**

There is no conflict of interest.